\documentstyle[psfig,prb,aps]{revtex}
\begin{document}

\headheight-1.0cm
\headsep0.0cm
\topmargin0.6cm
\textheight24cm

\draft

\twocolumn[\hsize\textwidth\columnwidth\hsize\csname
@twocolumnfalse\endcsname

\title{Density-functional theory study of the 
catalytic oxidation of CO over transition metal surfaces}
\author{C. Stampfl and M. Scheffler}
\address{Fritz-Haber-Institut der Max-Planck-Gesellschaft,
Faradayweg 4-6, D-14195 Berlin-Dahlem, Germany}
\maketitle

\begin{abstract}
In recent years due to improvements in calculation methods and increased 
computer power, it has become possible to perform first-principles
investigations for ``simple'' chemical reactions at
surfaces. We have carried out such studies for the
catalytic oxidation of CO at transition metal surfaces, in particular,
at the ruthenium surface for which unusual behavior compared
to other transition metal catalysts has been reported.
High gas pressure catalytic reactor experiments have revealed
that the reaction rate over Ru for oxidizing conditions
is the highest of the transition metals considered 
-- in contrast, under ultra high vacuum conditions,
the rate is by far the lowest.
We find that 
important for understanding the pressure dependence of the reaction
is the fact that  Ru\,(0001) can support
high concentrations of oxygen at the surface.
Under these conditions,
the O-metal bond is atypically weak compared to
that at lower coverages. 
We have investigated a number of possible reaction pathways for
CO oxidation for the conditions of 
high oxygen coverages, including
scattering reactions of {\em gas-phase}
CO at the oxygen covered surface (Eley-Rideal mechanism) 
as well as the 
Langmuir-Hinshelwood mechanism involving
reaction between {\em adsorbed} CO molecules and O atoms.
\end{abstract}

\vskip2pc]

\section{Introduction}
Carbon monoxide oxidation is one of the most extensively 
studied heterogeneous catalytic reactions. This  is due to both its
technological importance (e.g., in car exhaust catalytic converters
where the active components are transition metals such as
Pt, Pd, and Rh) and its ``simplicity''~\cite{king,oxygen,peden1}.
A {\em microscopic} understanding, however, 
of this most fundamental catalytic reaction is still lacking.
We note, however, that steps in this direction have recently been made via 
first-principles calculations~\cite{stampfl-co,alavi,eichler}.
Experimentally, it is  difficult to
probe the state of the reactants {\em during} reaction -- 
traditionally, information is only available before or after the event. 
Advances in surface science techniques, however, afford new 
information concerning the behavior of the reactants during
the reaction process. For example,
time-resolved scanning tunneling microscopy (STM)~\cite{stm}, 
time-resolved electron energy-loss spectroscopy (TREELS)~\cite{treels},
time-resolved infrared spectroscopy (TRIS)~\cite{tris}, and
(``fast'' high resolution)
 x-ray photoelectron spectroscopy (XPS)~\cite{photoemission}. 
On a larger scale are 
low energy electron microscopy (LEEM)~\cite{leem} 
and photoemission electron microscopy (PEEM)~\cite{peem}.
Theoretically,
a significant hindrance has been computational limitations,
but also, for a realistic description of certain reactions,
new theoretical developments
had to be awaited; for example, the 
generalized gradient approximation (GGA) for the exchange-correlation
functional has been shown to be
crucial for obtaining accurate activation barriers for 
hydrogen 
dissociation at metal and semiconductor surfaces~\cite{hammer1,hammer2}.

Past studies performed using surface science techniques, i.e.,
on well characterized single-crystal
metal surfaces under ultra high vacuum (UHV) conditions,
have shown that CO oxidation 
proceeds  via the Langmuir-Hinshelwood (L-H) mechanism in which 
reaction takes place between {\em chemisorbed} reagents~\cite{king}. 
Recent high gas-pressures catalytic reactor experiments, which afford
the study of chemical reactions under ``realistic'' high pressure and
temperature conditions, support the assignment of the L-H mechanism
for Pt, Pd, Rh, and Ir~\cite{peden1}.
For Ru, however, somewhat anomalous behavior was found which
indicated that reaction via scattering of CO molecules with
adsorbed O atoms may be taking place, i.e. via an Eley-Rideal-type
mechanism~\cite{peden3}. In particular, with pressures of about 10 torr and
for oxidizing conditions (i.e., at CO/O$_{2}$ pressure ratios $ < 1$)
the rate of CO$_{2}$ production
was found to be significantly higher than at the other
transition metal surfaces~\cite{peden3,peden2}; 
in contrast, 
under UHV conditions, the rate is extremely low over Ru\,(0001)~\cite{king,uhv1,uhv2}.
Unlike the other transition metals,
almost no chemisorbed CO could be
detected either during or after the reaction, and the 
kinetic data (activation energy and
pressure dependencies) was found to be markedly different to that of 
the other metals; in particular, highest rates occurred for high 
concentrations of oxygen at the surface, whereas
for the other metals, highest rates occurred for low O coverages.
Our studies show that Ru does behave differently to the other transition
metal catalysts in that high coverages of O can be supported on the
surface (up to desorption temperatures)
where the O-metal bond is significantly weaker than in the
lower coverage phases. Investigation of the energetics for CO$_{2}$
formation indicates that a Langmuir-Hinshelwood mechanism,
rather than an Eley-Rideal process is dominant.

\section{Calculation Method}
In order gain understanding into the apparently different behavior
of Ru for the CO oxidation reaction, and to obtain a microscopic picture
of this basic surface catalyzed reaction in general, we carried out 
density-functional theory (DFT) calculations~\cite{stumpf}.
We use the {\em ab initio} 
pseudopotential plane wave method and the supercell approach
where we employ the GGA~\cite{perdew}
for the exchange-correlation interaction.
We use {\em ab initio}, fully separable, norm-conserving
GGA pseudopotentials~\cite{troullier,martin}, where for the Ru atoms,
relativistic effects are taken into account using
weighted spin-averaged pseudopotentials.
The surface is modelled using a $(2 \times 2)$ surface unit cell 
with four layers of Ru\,(0001).
An energy cut-off  of 40 Ry is taken with three special
{\bf k}-points in the two-dimensional Brillouin zone~\cite{cunningham}.
The adsorbate structures are created on one side of the slab~\cite{stumpf}.
We relax the position of the atoms, keeping the
Ru atoms in the bottom two layers fixed at their bulk-like
positions.

\section{Oxygen on Ruthenium}
Under UHV conditions,
at room temperature, dissociative adsorption of O$_2$
results in an (apparent)  saturation coverage of $\Theta_{\rm O} \approx 1/2$,
corresponding to the formation of a $(2 \times 1)$-O structure~\cite{pfnur}.
At $\Theta_{\rm O}$=1/4, a $(2 \times 2)$-O phase forms~\cite{lindroos}. 
Here $\Theta_{\rm O}=1$ means that there are as many
O atoms as there are Ru atoms in the top layer.
In both
surface structures, O adsorbs in the hexagonal close-packed (hcp)
site.
Our earlier DFT-GGA calculations for O on Ru\,(0001)~\cite{stampfl,over}
indicated that
even higher coverage phases should form; 
namely,  a ($1 \times 1$)-O structure with coverage  $\Theta_{\rm O} = 1$,
as well as a $(2 \times 2)$-3O structure with coverage  $\Theta_{\rm O}
 = 3/4$.
As for the lower coverage structures,
 the O atoms occupy hcp sites.
The adsorption energy of O decreases notably with
increasing coverage and for the monolayer coverage phase,
the adsorption energy is
$\approx$0.7~eV less than that of the $(2 \times 2)$-O phase.
Both the $(2 \times 2)$-3O and $(1 \times 1)$-O 
structures have subsequently been created experimentally with the use
of NO$_{2}$ or high gas pressures of O$_{2}$~\cite{over,kostov-3o,over2,menzel},
and the atomic structure 
verified by  dynamical low-energy electron diffraction (LEED) intensity
analyses.
NO$_{2}$ readily dissociates at elevated temperatures in the presence of
adsorbed oxygen delivering atomic oxygen to the surface while NO desorbs.
Furthermore, it has been demonstrated that {\em after} completion
of the monolayer  oxygen structure,
additional oxygen can enter the subsurface region~\cite{over,mitchell}
at elevated temperatures.
Formation of the higher coverage phases ($\Theta=$3/4 and 1.0)
from gas-phase O$_{2}$ at ``usual'' exposures under UHV conditions
is apparently kinetically 
hindered  by activation barriers for O$_{2}$ dissociation, induced by
the pre-adsorbed oxygen atoms at coverage $\Theta_{\rm O} \approx$0.5.
With respect to the high pressure catalytic reactor experiments mentioned
above,
because the conditions under which the highest rates of CO$_{2}$
formation were reported involved high O$_{2}$ 
partial gas pressures and oxidizing conditions,
there will be a  significant attempt frequency of O$_{2}$
to overcome activation barriers for dissociative adsorption.
Thus, it is likely that during reaction,
the oxygen coverage on the surface approaches one
monolayer.
We note also that in the catalytic reactor experiments it may be unlikely
that there is significant subsurface oxygen present since the temperature
range  studied in the experiment 
of 380~K to 500~K is less than that of 600~K 
at which oxygen is reported
to enter the subsurface region
with an appreciable rate (using NO$_{2}$)~\cite{over,mitchell} and
auger electron spectroscopy indicated a coverage of about one monolayer.

\section{Reaction via gas-phase CO with adsorbed O}

In earlier publications we reported our
investigation for reaction via scattering of
gas-phase CO with adsorbed O~\cite{stampfl-co} so here we only briefly describe 
the results.

The $(1 \times1 )$-O phase 
is assumed to cover the whole surface and we investigate the interaction of CO 
with this oxygen-covered surface.
For a given lateral position,
CO is placed well above the surface (with the
C-end down) and the total  energy calculated for decreasing
distances of CO from the surface. 
All atomic positions are relaxed except that of the C atom which is held
fixed (and the bottom two Ru layers).
%
%
\begin{figure}
\vspace{-10mm}
\psfig{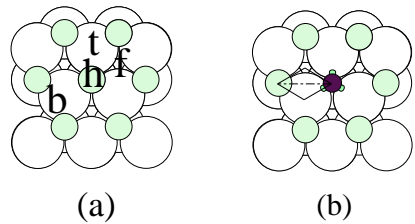}
\caption{(a) Schematic diagram illustrating the on-top (t), fcc (f),
hcp (h), and
bridge (b) sites of $(1 \times 1)$-O/Ru\,(0001). Small and large
circles represent O and Ru atoms, respectively.
(b) Top view of CO adsorbed in an O-vacancy
of $(1 \times 1)$-O/Ru\,(0001). The full lines represent the
(symmetry inequivalent) region within which we consider reaction takes  place.
CO is indicated by the small black circle.}
\end{figure}
We considered a number of lateral positions: the
on-top and  fcc sites, with respect to the Ru\,(0001) substrate,
a bridge site between two adsorbed O atoms, as well as 
directly above an adsorbed O(a) atom (see Fig.~1a).
We find that an energy barrier 
begins to build up at about 2.5~\AA\, from the surface (with respect
to the average position of the O atoms) for
all sites, reflecting 
a {\em repulsive} interaction of CO with the O-covered surface.
A very weak physisorption well of about 0.04~eV is also found
above the O-adlayer.
Its position is $\approx$3.0~\AA\, from the surface
for the on-top and fcc sites, and at about
$\approx$3.5~\AA\, for the bridge site and the approach directly
over the O atom. 
Thus, for a full monolayer coverage of oxygen on the surface,
CO is unable to form a chemical bond with the substrate and
the L-H reaction mechanism is therefore prevented.

In order to obtain a more detailed understanding of
CO$_{2}$ formation 
via  a scattering reaction, we 
evaluated an appropriate cut through the high-dimensional
potential energy surface (PES) (see Ref.~\cite{stampfl-co}). This
cut is defined by two variables: the
vertical position of the C atom
and the vertical position of the reacting O(a) adatom directly
below.
Initially, with the CO-axis is held perpendicular to the surface
we find that the activation barrier
for CO$_2$ formation is  $\approx$1.6~eV.
When the tilt angle of the CO-axis
is allowed to relax, 
the energy barrier is reduced to about 1.1~eV.
The transition state (depicted in the inset of Fig.~2) has a 
%
%
\begin{figure}
\vspace{-10mm}
\psfig{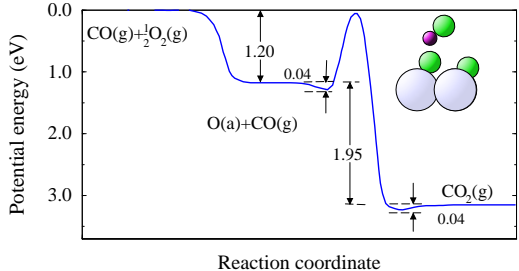}
\caption{Calculated energy diagram for the E-R mechanism of CO oxidation
at Ru\,(0001). Note that the depths of the physisorption wells are
exaggerated for clarity.
The transition state geometry is indicated in the inset.
The large, medium, and
small (dark) circles represent Ru, O, and C atoms, respectively.}
\end{figure}
``bond angle'' of $\approx 131^{\circ}$ and the reacting
O(a) atom is $\approx$0.35~\AA\, above the other O atoms in the surface
unit cell. The C-O(a) bond length is 1.50~\AA\, (stretched by 27~\% compared to
the calculated bond length of a free CO$_{2}$ molecule which 
is 1.18~\AA\,; the experimental value is 1.16~\AA\,~\cite{pople}), and
the bond length of CO is 1.17~\AA\, (the calculated value 
of a free CO molecule is 1.15~\AA\,; the experimental value
is also 1.15~\AA\,~\cite{pople}).
After onset of reaction, the CO-O(a) bond begins to develop and the
O(a)-Ru bond is weakened. It  then becomes
energetically unfavorable for the reacting complex to be at the surface
and it is strongly repelled towards the vacuum region.
The resulting energy diagram is shown in Fig.~2.
It can be seen that there is a significant energy gain from the surface 
reaction of 1.95~eV.

Using the determined activation barrier in 
a simple Arrhenius-type equation with the prefactor obtained from
consideration of
the number of CO molecules hitting the surface per site per second 
at a given temperature and pressure,
we can estimate the reaction rate~\cite{stampfl-co}. 
This will give an upper bound since for other orientations of the molecule
the barrier is larger.
The rate is found to be
significantly lower
(by $3 \times 10^{-6}$) than that measured experimentally~\cite{peden2}.
This indicates that this mechanism
alone cannot explain the enhanced CO$_{2}$
turnover frequency as was speculated.
To investigate other  possible reaction channels, we consider it
conceivable that there are vacancies in the perfect
$(1 \times 1)$-O adlayer (see Ref.~\cite{stampfl-co} for an estimate
of the O-vacancy density).
CO molecules may then adsorb at these vacant sites and react via
a L-H mechanism.  In the following section we 
investigate such a L-H reaction mechanism.

\section{Reaction between adsorbed CO and O}

We first consider the energetics of adsorption of CO into a vacant
hcp site of the monolayer oxygen structure. Interestingly, we
find  that there is an activation barrier of $\approx$0.3~eV. 
We expect, however, that this barrier will relatively 
easily be overcome at the high gas pressures used in 
the catalytic reactor experiments.
From Fig.~1b it can be seen that CO is 
closely surrounded by a hexagonal arrangement of  six O atoms. 
Even though the reactants are 
already very close, there is still a repulsive rather
than an attractive interaction between
CO and the neighboring O atoms.
In determining the reaction path we
consider reaction in the
inequivalent area of the surface, 
indicated in Fig.~1b by the continuous lines.
There are clearly a number of possible ways that the reaction between CO and
a neighboring O atom could proceed:
for example, the CO molecule may approach an O atom, an O atom
may approach the CO molecule, or the reactants may both move towards 
each other over one Ru atom. For each of these scenarios there are
obviously also a number of possible
reaction paths (i.e., via the on-top site, bridge and fcc site, etc).  
To determine the minimum energy pathway, energy barrier, and
associated transition state, is clearly not a simple problem.
Recent {\em ab initio} studies of surface reactions and
dissociation of diatomic molecules
at surfaces have attempted ``direct methods'' for finding
the lowest energy reaction pathway~\cite{eichler,norskov}.
In the present work, however, 
we use the ``standard grid approach'' of constructing
various relevant PESs since we are also interested in the {\em shape}
of the PES  away from the minimum energy reaction pathway.

In view of the weaker CO-metal
bond strength compared to that of the O-metal bond strength at this
coverage, i.e. 0.85~eV compared to 2.09~eV (with respect to gas
phase 1/2 O$_{2}$), we first consider
reaction via movement of CO towards the O atom. 
This in fact turns out to have the lowest energy
pathway of the ones we considered with a barrier of $\approx$1.51~eV.
This value is consistent with the recent experimental estimate of
$>$1.4~eV for the case of high oxygen coverages on the 
surface~\cite{boettcher}.
Other possible reaction pathways considered were for O to move
towards CO and for O and CO to move towards each other.
The lowest energy pathway found for these scenarios were
at least 0.2 and 0.3~eV higher, respectively.
We point out that in this work we have only considered the 
most obvious reaction paths, and in order
to explore in more detail the very complex nature of this high-dimensional
potential energy surface, further calculations are required.

As a first step we investigate the energetics
for different fixed lateral positions of the C atom within the area shown
in Fig.~1b.
We initially keep the lateral position of the O atoms fixed, but allow 
vertical relaxations.
For CO moving towards the on-top site (see Fig.~1), a strong repulsion between 
C and the two symmetrically equivalent O atoms develops giving rise
to a large energy barrier of 2.53~eV.
Thus, the pathway over the on-top site is energetically unfavorable.

Interestingly, the situation is somewhat different for CO moving towards
the fcc site: in this direction there is also
the build up of an energy barrier; 
on overcoming the barrier, however, 
there is an {\em attractive} interaction between C and the
{\em two} symmetrically  equivalent O atoms (see Fig.~1b).
When the O atoms are then allowed to laterally relax, we find
the formation of a carbonate-like species.
The atomic geometry  is
depicted in Fig.~3a. We note that in our investigations of the CO-gas
%
%
\begin{figure}
\vspace{-10mm}
\psfig{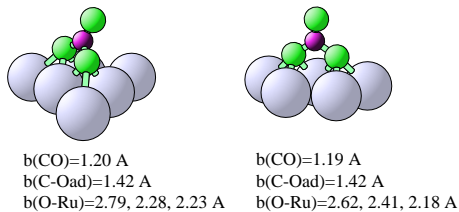}
\caption{Atomic geometry of identified carbonate species at
(a) a vacancy in the monolayer oxygen structure, and (b) at
the perfect $(1 \times 1)$-O/Ru\,(0001) surface.
The large, small, and
small dark  circles represent Ru, O, and C atoms, respectively.
The various bond lengths are indicated (in \AA).}
\end{figure}
scattering reaction, we also identified the stability
of such a carbonate species
on the fully O-covered surface;
the atomic geometry is similar as can be seen from Fig.~3b.
Formation of this species was also found to involve
a significant energy barrier.
We note that 
experimental identification of carbonate species in CO oxidation reactions 
over other transition metals have
been reported (e.g., Ref.~\cite{carbonate}), where they
may possibly act as an intermediary. The actual
role they play in the carbon monoxide oxidation
reaction for this system, however, is at present unknown.

%
%
\begin{figure}
\vspace{-10mm}
\psfig{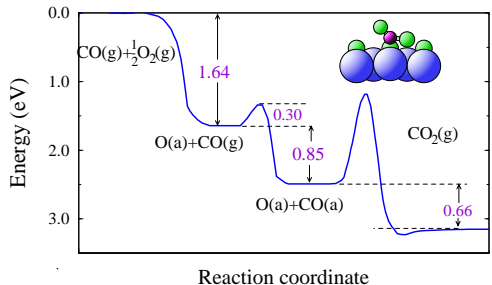}
\caption{Calculated energy diagram for the L-H mechanism of CO oxidation
at Ru\,(0001).
The transition state geometry is indicated in the inset.
The large, small, and
small darker circles represent Ru, O, and C atoms, respectively.}
\end{figure}
We find that the minimum energy pathway found for CO$_{2}$
formation corresponds to one where 
the CO molecule moves essentially directly 
towards the O atom. The associated energetics
are shown in Fig.~4 where we have constructed the energy
diagram~\cite{comment}. 
Similarly to our study of the E-R mechanism, we also find a small physisorption
well for CO$_{2}$ above the surface.
The corresponding transition state geometry is depicted in the inset of
Fig.~4.
In this geometry, the C-O(a) 
bond is almost parallel to the surface and
the CO axis is bent away from O(a) yielding a bent
CO-O(a) complex with a bond angle of 125$^{\circ}$; similar
to that found for the E-R mechanism. The  C-O(a) bond length is 
1.59~\AA\, (about 35~\% stretched compared to that in CO$_{2}$)
and the  CO bond length is 1.18~\AA\,.
%
%
\begin{figure}
\vspace{-10mm}
\psfig{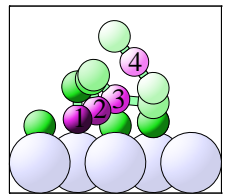}
\caption{Atomic positions a selected points
along the reaction energy pathway CO$_{2}$
formation.
The large, small, and
small darker  circles represent Ru, O, and C atoms, respectively.}
\end{figure}
Figure~5 shows the atomic geometry at selected positions along the reaction path
to CO$_{2}$ formation.
Initially CO is in the vacancy and O(a) in the hcp site. As CO approaches
O(a), repulsive forces build up on both the C and O(a) atoms and the molecular
axis of CO begins to tilt away from O(a). 
At the transition state,
CO and O begin to lift off the
surface as they break their metal bonds in favor of developing
a C-O(a) bond. Interestingly, the distance that O moves away from
the surface by is $\approx$0.36~\AA\,; very similar to that found
in the E-R mechanism which was $\approx$0.35~\AA\,. 
As CO$_{2}$ begins to form, the molecular axis quickly straightens out
to its linear geometry.

The corresponding valence
electron density and density difference distributions are shown in Fig.~6.
The latter is constructed by subtracting from the electron density of the
CO,O/Ru\,(0001) system, those of the O-covered surface and a free CO molecule.
From the electron density difference distributions, with respect to the starting
configuration of CO in the vacancy, it can be seen that 
as CO moves towards O(a), firstly there is less electron density in
the CO $2\pi^{*}$ orbitals, indicating a weakening of the C-metal bond.
Also, electron density has been depleted from the O(a) orbital
pointing towards the CO molecule and an increase occurs into 
orbitals in the orthogonal direction (pointing in a near-perpendicular
direction to the surface).
The redistribution is an effect of Pauli repulsion.
At the transition state, the onset of bond formation can be seen
between these latter O(a) orbitals and the $2\pi^{*}$-like orbitals of the
C atom.
We can notice 
that CO is bonded only weakly to the metal through a $2\pi^{*}$-like orbital.
The bond of the reacting O(a) to 
the surface is also significantly weakened at the transition state as
can be see from the lower panel showing the total valence
electron density. 
Interestingly, the density difference plot for the transition state
is very similar to that of the E-R process (see Fig.~7 of 
Ref.~\cite{stampfl-co}).
In the last panel (leftmost),
significant accumulation of electron density can be seen
between the C and O(a) atoms 
as the CO$_{2}$ molecule is practically formed.

To summarize,
our studies indicate that a Langmuir-Hinshelwood mechanism rather
than an Eley-Rideal mechanism is the {\em dominant} reaction process giving
rise to the reported increase in reactivity of
ruthenium for the CO oxidation reaction as measured in the
high pressure catalytic reactor experiments as compared to under
UHV conditions.
We find that for high coverages of O on the surface,
which are attainable under the sufficiently high oxygen pressures
used in the experiment, the O adsorption energy is notably 
weaker than the lower coverage phases that form under UHV conditions.
Furthermore, the adsorption energy of  CO in the presence of
high O coverages is weaker than in the presence of low O coverages.
These factors, together with the 
close proximity of the reactants
is thought to give rise to the observed increase in reactivity of Ru
and the reported anomalous behavior on partial gas pressure.

Finally, we like to mention 
very recent results of CO oxidation experiments:
For the case of very high concentrations 
of oxygen at the surface
(one monolayer on the surface plus oxygen occupying {\em subsurface} sites)
which can be prepared after formation of the $(1 \times 1)$ phase
by using either NO$_{2}$ or high gas pressures of O$_{2}$ 
at {\em elevated} temperatures, reaction rates
notably greater than that from the on-surface
monolayer oxygen structure have been measured~\cite{boettcher,boettcher2}.
These high rates have been proposed to be connected to the existence of
copious amounts of oxygen in the
subsurface region.
These are clearly very interesting results and require more detailed
investigations
in order to understand this behavior.
%
%
\begin{figure}
\vspace{-10mm}
\psfig{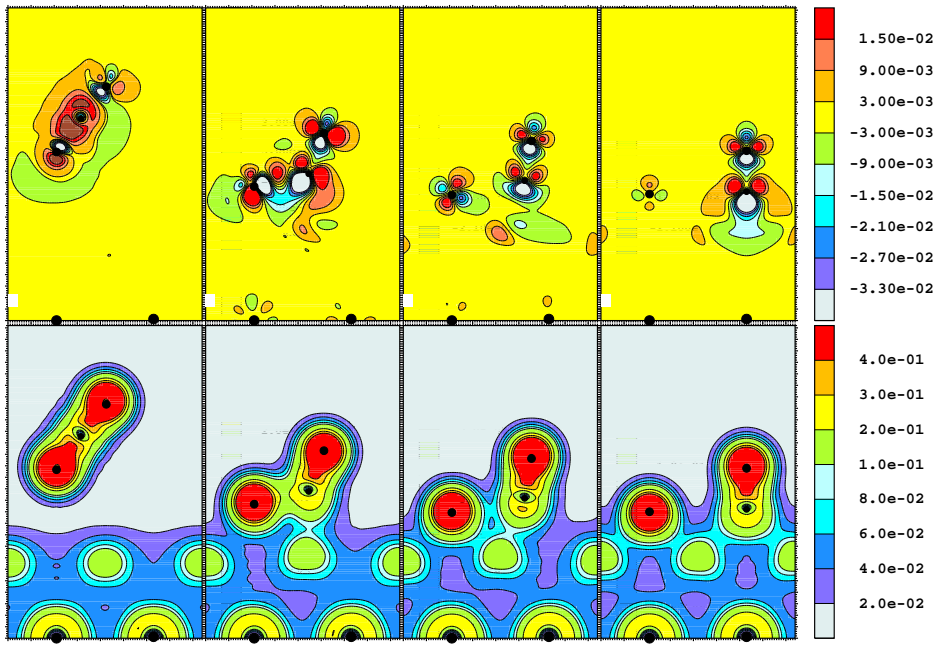}
\caption{Valence electron density (lower panels)
and density difference distributions (upper panels)
for selected positions along the reaction path. The units
are in $e$ bohr$^{-3}$. For the upper leftmost panel the contours differ:
The first negative contour is at $-3 \times 10^{-3}$
and the spacing is $-15 \times 10^{-3}$ and the
first positive contour is at $3 \times 10^{-3}$
and the spacing is $30 \times 10^{-3}$.}
\end{figure}

\subsection{Conclusion}
We have performed density functional theory calculations in order
to investigate the catalytic oxidation of carbon monoxide over
Ru\,(0001) for the conditions of high oxygen coverages on the
surface for which highest reaction rates of CO$_{2}$ formation
have been reported.
It is only by exposure of the surface to high O$_{2}$ gas pressures, or with the use
of strongly oxidative molecules such as NO$_{2}$
 that high oxygen coverages can be achieved.
In this case the O-metal bond strength is notably weaker compared to the
lower coverage structures that form under UHV conditions, and as such,
is expected to be more reactive. In the 
present work we concentrated mainly on the microscopic description
and study of the Langmuir-Hinshelwood reaction mechanism. We identified
a bent transition state for CO movement towards an adsorbed O atom
with an associated energy barrier of approximately 1.5~eV.
In this configuration the CO and the adsorbed O atom have substantially
weakened (and significantly stretched) their bond to the substrate.


\begin{references}
\bibitem{comment} We point out that allowing lateral atomic relaxations
of the neighboring O atoms during determination of the reaction pathway may
lead to a slightly lower energy barrier for CO$_{2}$ formation.
\end{references}
\end{document}